\documentclass[twocolumn,showpacs,superscriptaddress,amsmath,amssymb]{revtex4}
\usepackage{graphicx}
\usepackage{color}
\usepackage{dcolumn}
\usepackage{bm}

\newcommand{\ignore}[1]{}
\newcommand{\B}{{\mathcal B}}

\begin{document}

\title{Boolean networks with veto functions}
\author{Haleh Ebadi}
\affiliation{Bioinformatics, Institute for Computer Science, Leipzig University,
H\"{a}rtelstrasse 16-18, 04107 Leipzig, Germany
}

\author{Konstantin Klemm}
\affiliation{Bioinformatics, Institute for Computer Science, Leipzig University,
H\"{a}rtelstrasse 16-18, 04107 Leipzig, Germany
}
\affiliation{Bioinformatics and Computational Biology, University of Vienna,
W\"{a}hringerstra{\ss}e 29, 1090 Vienna, Austria}

\affiliation{Theoretical Chemistry, University of Vienna,
W\"{a}hringerstra{\ss}e 17, 1090 Vienna, Austria}

\affiliation{School of Science and Technology, Nazarbayev University,
010000 Astana, Kazakhstan}
\date{\today}

\begin{abstract}
Boolean networks are discrete dynamical systems for modeling regulation
and signaling in living cells. We investigate a particular class of
Boolean functions with inhibiting inputs exerting a veto (forced zero) on
the output. We give analytical expressions for the sensitivity of these
functions and provide evidence for their role in natural systems. In an
intracellular signal transduction network [Helikar et al., PNAS (2008)],
the functions with veto are over-represented by a factor exceeding the
over-representation of threshold functions and canalyzing functions in the
same system. In Boolean networks for control of the yeast cell cycle
[Fangting Li et al., PNAS (2004), Davidich et al., PLoS One (2009)], none
or minimal changes to the wiring diagrams are necessary to formulate their
dynamics in terms of the veto functions introduced here.
\end{abstract}


% 89.75.Fb Structures and organization in complex systems
% 87.16.Yc Regulatory genetic and chemical networks
% 05.45.-a Nonlinear dynamics and chaos
% 45.05.+x General theory of classical mechanics of discrete systems

\maketitle

\section{Introduction}

Networks of chemical interactions are responsible for the signalling and control
in all living systems, from the unicellular bacteria to large multicellular
organisms \cite{Alon:2006}. We are witnessing a rapid
increase of biochemical measurements. These results need to be complemented by
appropriate models in order to elucidate common principles of such systems and
generate predictions testable by further experiments. A variety of modeling
approaches exist, ranging from the chemical master equation or stochastic
simulation of reactions for a few types of molecules to purely qualitative
wiring diagrams that summarize existing interactions \cite{deJong:2002}.

A particularly successful approach of simplification for control networks of up
to hundreds of nodes is the discretization of chemical signals into on/off
states evolving in discrete time \cite{Kauffman:1969,Bornholdt:2005}. These
kinds of models, called Boolean networks, are formally equivalent to circuits of
digital electronics with logical gates.

For more and more control systems, the essential time course and response to
perturbations are accurately reproduced by a dedicated Boolean network
\cite{Kauffman:2003,Albert:2003,Li:2004,Helikar:2008}. Such system-specific
Boolean models are obtained from known interactions in the literature
\cite{Davidich:2008}, by discretizing existing models of differential equations
\cite{Davidich:2008b} or by direct inference from high-throughput experimental
data \cite{Wang:2012,Saadatpour:2013}.

Long before the data-driven definition and refinement of system-specific
networks, however, statistical ensembles of Boolean networks were studied,
seeking generic properties of these discrete dynamical systems
\cite{Kauffman:1969,Derrida:1986}. There the Boolean functions are assigned to
the nodes randomly over the set of all functions with a given
number of inputs. When increasing the average number of interactions, such
random Boolean networks display a transition from ordered behaviour dominated by
fixed points to ``chaotic'' dynamics with transients and periodic
attractors of length exponential in system size \cite{drossel-review}.

Random Boolean networks may now serve as null models in comparison to
system-specific models. One way of refinement of these null models is by
restricting the set of Boolean functions to realistic ones. Though the
repertoire of combinatorial biochemical interactions, e.g.\ between
transcription factors and binding sites \cite{Buchler:2003}, enables
construction of complicated logical functions, relatively simple truth tables
abound in real systems. One class of naturally occurring input-output
relations are canalyzing \cite{Harris:2002}: a certain truth
value at one argument fully determines the output of the Boolean function.
Using nested canalyzing functions, where the residual function after removal
of one canalyzing input is again canalyzing, the dynamics of the Boolean
networks is ensured to be non-chaotic \cite{Kauffman:2004}.

Beyond making null models more realistic, the usage of a specifically restricted
set of Boolean functions also offers advantages in the numerical treatment of
Boolean dynamics and, in particular, the evolution of the networks
\cite{Bornholdt:1998}. Such
simulation scenarios frequently use threshold functions \cite{Rohlf:2002}, whose
output is active only if a weighted sum of the inputs exceeds a certain value.
Similar to $\pm J$-spin glasses \cite{Binder:1986} but keeping couplings
asymmetric in general, these threshold functions employ binary weights taking
values +1 (activation), -1 (inhibition), and an entry 0 representing
absence of a coupling in the interaction matrix. For a function with $k$
inputs, this choice reduces the set of available functions from
$2^{2^k}$ to at most $3^k$. Threshold functions, however, are not the only
practical choice of Boolean functions where inputs are assigned binary labels
in this manner.

Here we investigate a class of functions with strong inhibition, which we call
veto functions. As is the case with threshold functions, inputs have binary
labels, activating or inhibiting. However, the output is shut off by a single
inhibitory signal regardless of other inputs. We calculate the sensitivity of
these functions  and provide two instances of relevance for biological systems.
First, veto functions preferentially occur  in a large Boolean network of
inter-cellular signalling. Second, known wiring diagrams for control of 
yeast cell cycles generate the correct trajectories under veto functions.

%%%%%%%%%%%%%%%%%%%%%%%%%%%%%%%%%%%%%%%%%%%%%%%%%%%%%%%%%%%%%%%%%%%%%
\section{Definitions and notation} \label{sec:def}
%%%%%%%%%%%%%%%%%%%%%%%%%%%%%%%%%%%%%%%%%%%%%%%%%%%%%%%%%%%%%%%%%%%%%

A Boolean function is a mapping
\begin{equation}
f:\{0,1\}^k \rightarrow \{0,1\}
\end{equation}
of $k$ binary valued inputs with a single binary output. The number of
inputs $k$ is called the {\em arity} of $f$ \cite{Burris:1981,wikipedia:arity}.
For testing the sensitivity of $f$ under changes of the state of one input,
we define the negation (``flip'') of
the $i$-th component on a Boolean vector $\sigma \in \{0,1\}^k$ as the vector
$\sigma^{\updownarrow i}$ with
\begin{equation}\label{eq:negation}
(\sigma^{\updownarrow i})_j \neq \sigma_j \Leftrightarrow i=j ~.
\end{equation}
Not all $k$-ary functions actually depend on all $k$ inputs. We call input $j$
of function $f$ {\em spurious} if 
\begin{equation}
f(\sigma) = f(\sigma^{\updownarrow j})
\end{equation}
for all input vectors $\sigma$. Thus input $j$ is spurious if $f$ can be computed
without knowing the value at input $j$.

A Boolean network is a time- and state-discrete dynamical system given by an iteration
\begin{equation}\label{eq:boolmap}
\sigma(t+1) = F(\sigma(t))
\end{equation}
on a time-dependent binary state vector $\sigma \in \{0,1\}^N$.
The map
\begin{equation}
F:\{0,1\}^N \rightarrow \{0,1\}^N
\end{equation}
is a collection of $N$ Boolean functions $f_1,f_2,\dots,f_N$ each of
arity $N$. Note that $F$ maps Boolean vectors to Boolean vectors and thus
can be iterated. A Boolean function $f$  has a single value 0 or 1 as output.
In practical and realistic scenarios, the functions depend only on a
small subset of all $N$ inputs, all other inputs are spurious.
These systems are then characterized by their
sparse interaction networks and hence the name Boolean network. See
section~\ref{sec:cellcyc} for examples.

In alternative notation \cite{Aldana:2003}, a Boolean network is an $N$-tuple of
Boolean functions $(f_1,f_2,\dots,f_N)$ where function $f_i$ has an arity
$k(i) \le N$ not necessarily the same for all $i \in\{1,2,\dots,N\}$. Coupling
between nodes is encoded by assigning each node $i$ a $k(i)$-tuple
$(j_{i,1},j_{i,2},\dots,j_{i,k(i)})$ listing the indices of the nodes feeding
into node $i$. Node $i$ updates its state as
\begin{equation}
\sigma_i(t+1) = f_i(\sigma_{j_{i,1}}(t),\sigma_{j_{i,2}}(t),\dots,\sigma_{j_{i,k(i)}}(t))~.
\end{equation}
This notation and the one by Eq.~(\ref{eq:boolmap}) are equivalent because they
describe the same set of systems. A vector mapping $F$ according
to Eq.~(\ref{eq:boolmap}) is more compact and in line with the standard
notation of multi-dimensional dynamical systems (flows and iterated maps) as
functions on vector spaces. Not all $N \times N$ possible
couplings between the $N$ state variables need to be present. Therefore, we
need to explicitly deal with Boolean functions having spurious inputs.

%%%%%%%%%%%%%%%%%%%%%%%%%%%%%%%%%%%%%%%%%%%%%%%%%%%%%%%%%%%%%%%%%%%%%
\section{Canalyzing and threshold functions} \label{sec:can_thr}
%%%%%%%%%%%%%%%%%%%%%%%%%%%%%%%%%%%%%%%%%%%%%%%%%%%%%%%%%%%%%%%%%%%%%

Canalyzing functions, sometimes called forcing functions, have been studied
widely in the context of Boolean networks \cite{Kauffman:2003,Harris:2002}. Canalyzation means that a certain
value at one of the inputs determines the output, regardless of the other
inputs. For a Boolean function $f$, the input with index $j$ is canalyzing 
if there are $b,c \in \{0,1\}$ such that for all $\sigma \in \{0,1\}^k$
\begin{equation}
\sigma_j=b \Rightarrow f(\sigma)=c~.
\end{equation}
Then $b$ is the canalyzing value and $c$ is the canalyzed value. A Boolean function
$f$ is called canalyzing if $f$ has a canalyzing input.

A different widely used class of functions are those defined by a weight vector and
a threshold \cite{Li:2004,Lau:2007,Davidich:2008}.
A $k$-ary Boolean function $f$ is a {\em general threshold function} if there is
a weight vector $w=(w_1,w_2,\dots,w_k) \in \mathbb{R}^k$ and a threshold
$\theta \in \mathbb{R}$ such that
\begin{equation} \label{eq:thrgeneral}
f(\sigma) = H ( \sum_{j=1}^k w_j \sigma_j - \theta) 
\end{equation}
for all $\sigma \in \{0,1\}^k$, using the step function
$H: \mathbb{R} \rightarrow\{0,1\}$ with
$H(x) = 1$ if and only if $x > 0$.

Here we consider the restriction to the case of discrete weights $w_j \in \{-1,0,+1\}$ for all inputs
$j$ and a vanishing threshold $\theta=0$. See the recent work by Rybarsch and Bornholdt 
\cite{Rybarsch:2012} for
a motivation of this choice in the context of biochemical regulation. By
{\em threshold function}, we denote a member of this restricted set of functions.

%%%%%%%%%%%%%%%%%%%%%%%%%%%%%%%%%%%%%%%%%%%%%%%%%%%%%%%%%%%%%%%%%%%%%
\section{Veto functions} \label{sec:veto}
%%%%%%%%%%%%%%%%%%%%%%%%%%%%%%%%%%%%%%%%%%%%%%%%%%%%%%%%%%%%%%%%%%%%%

For veto functions, similar to threshold functions, the set of inputs is
divided into subsets of activating, inhibitory and irrelevant inputs. The
output of a veto function is active if and only if all inhibitors are off and
at least one activator is on. Formally, a $k$-ary Boolean function $f$ is a
veto function, if there are
$A,I \subseteq \{1,\dots,k\}$ with $A \cap I = \emptyset$ such that
for all $\sigma \in \{0,1\}^k$,
\begin{equation}\label{eq:veto}
f_{v}(\sigma) = 1 \Leftrightarrow \forall j\in I: \sigma_j = 0 \text{ and } 
\exists l \in A: \sigma_l=1 
\end{equation}
Equivalently, veto functions may be defined by restricting the set of
general threshold functions. Then $f$ is a veto function if there is a
weight vector $w \in \{-k,0,+1\}^k$  such that  Eq.~(\ref{eq:thrgeneral}) holds
for all state vectors $\sigma \in \{0,1\}^k$ and threshold $\theta=0$. The
choice of $-k$ as
the weight of an inhibitor keeps the sum below the threshold irrespective of
activating inputs.

%%%%%%%%%%%%%%%%%%%%%%%%%%%%%%%%%%%%%%%%%%%%%%%%%%%%%%%%%%%%%%%%%%%%%
\section{Counting functions that depend on all their inputs}
\label{sec:counting}
%%%%%%%%%%%%%%%%%%%%%%%%%%%%%%%%%%%%%%%%%%%%%%%%%%%%%%%%%%%%%%%%%%%%%

For the data analysis in the following section, further notation and 
considerations are required for the counting of Boolean functions without
spurious inputs. Spurious inputs are absent in empirical data of networks, where
each input of a node represents a real regulatory interaction that does influence
the output. In order to assess if functions with a certain property are over- or
under-represented in real data, a reasonable null model is to be based only on
functions that depend on all inputs.

We shall see that threshold and veto functions without spurious inputs are easy
to identify and count due to the parameterization by a weight vector. This
simplicity is lacking in other classes of functions, in particular all functions
(unrestricted class) and canalyzing functions. We provide recursions for counting
Boolean functions without spurious inputs to cope with these classes. In order
to provide methodology for general types of Boolean functions in
future work, we introduce detailed mathematical formalism as follows.

By $\B$ we denote the set of all Boolean functions on finitely many inputs;
for $k\in \mathbb{N}\cup\{0\}$, we call $\B_k$ the set of all $k$-ary Boolean
functions, so
\begin{equation}
\B = \bigcup_{k=0}^\infty \B_k~.
\end{equation}
We denote the restriction of $\B$ to functions with a given property $\pi$ as
$\B^{(\pi)}$, and the further restriction to $k$ inputs as $\B^{(\pi)}_k$. Here
we are concerned with the three properties  canalizing, threshold and veto, so
$\pi \in \{\text{can},\text{thr},\text{veto}\}$ and the corresponding function
sets $\B^\text{can}$, $\B^\text{thr}$, $\B^\text{veto}$. By $\B^\ast$ we denote
the restriction of $\B$ to functions without spurious inputs; applying the same
restriction to $\B_k$ and $\B^{(\pi)}_k$, we use the symbols $\B^\ast_k$ and
$\B^{\ast,(\pi)}_k$.

For a threshold function or a veto function, the situation is quite simple. Each
zero entry in a representing weight vector $w$ is a spurious input. So let us
consider only functions with weight vectors $w \in \{-1,+1\}^k$ (for threshold
functions) or $w \in \{-k,+1\}^k$ (veto functions). For these, the weight vector
is unique and a single activating input ($j$ with $w_j=+1$) renders all inputs
non-spurious. Therefore, each combination of admissible non-zero weights, except
for the all-negative weight vector, represents a function in
$\B_k^{\ast,\text{thr}}$ and $\B_k^{\ast,\text{thr}}$, so we obtain
\begin{equation}
|\B^{\ast,\text{veto}}_k| = |\B^{\ast,\text{thr}}_k| = 2^k -1~.
\end{equation}

This straight-forward combinatorics is not the general case. We do not
find a representation in terms of weight vectors for each class of Boolean functions.
We now establish insight for a set of Boolean functions with a property $\pi$ that is
closed under permutations of inputs and removal / addition of spurious inputs. This
assumption is fulfilled for unrestricted Boolean functions as well as
the three properties $\{\text{can},\text{thr},\text{veto}\}$. The number of functions with
property $\pi$ without spurious inputs, is obtained recursively as
\begin{equation} \label{eq:unspur_iteration}
|\B^{\ast,(\pi)}_k| = |\B^{(\pi)}_k| - \sum_{j=0}^{k-1}  {k \choose j}  |B^{\ast,(\pi)}_j|~.
\end{equation}
For each $k$-ary function with exactly $k-j$ spurious inputs, these may be removed to arrive
at a unique $j$-ary function. The multiplicity of such $k$-ary functions reducing to
the same $j$-ary function in this way is given by the binomial factor, counting the
combinations in which spurious and non-spurious inputs are assigned.

A Boolean function with $k$ inputs takes $2^k$ different input vectors, to each of
which an output value is assigned independently. Thus there are $|\B_k| =
2^{2^k}$ Boolean functions of arity $k$. By inserting this result into 
Eq.~(\ref{eq:unspur_iteration}), the number of Boolean functions without spurious
inputs is obtained. For canalyzing functions, $|\B^{\ast,\text{can}}_k|$ is calculated
by the same equation using the results $|\B^{\text{can}}_k|$ from Just and co-authors
\cite{Just:2004}.

%%%%%%%%%%%%%%%%%%%%%%%%%%%%%%%%%%%%%%%%%%%%%%%%%%%%%%%%%%%%%%%%%%%%%
\section{Over-representation of functions in a signaling network} \label{sec:over}
%%%%%%%%%%%%%%%%%%%%%%%%%%%%%%%%%%%%%%%%%%%%%%%%%%%%%%%%%%%%%%%%%%%%%

\begin{table}
\caption{\label{tab:raw}
Counts  in the data set. The column $a_k$ is for the total count of
$k$-ary functions, the following three columns count veto, canalyzing and
threshold functions. The last line gives the summation over all arities.
}
\centering \setlength{\tabcolsep}{8pt}
\begin{tabular}{|r|rrrr|} 
\hline\hline
$k$ & $a_k$ & $a_k^\text{veto}$ & $a_k^\text{can}$ & $a_k^\text{thr}$ \\ [0.5ex] 
\hline 
1 & 27 & 27 & 27 & 27\\ 
2 & 23 & 12 & 21 & 12\\
3 & 21 & 5 & 18 & 2\\
4 & 29 & 4 & 15 & 2\\
5 & 11 & 1 & 4 & 1\\
6 & 10 & 0 & 5 & 0\\
7 & 8 & 0 & 3 & 0\\
8 & 10 & 0 & 4 & 0\\
9 & 5 & 0 & 3 & 0\\
10 & 5 & 0 & 5 & 0\\
11 & 1 & 0 & 1 & 0\\
12 & 1 & 0 & 0 & 0\\
13 & 0 & 0 & 0 & 0\\
14 & 1 & 0 & 1 & 0\\
\hline  $\sum_k$ & 152 & 49 & 107 & 44\\
  \hline
\end{tabular}
\end{table}

\begin{table}
\caption{\label{tab:over}
Over-representation of the three types of functions 
} 
\centering \setlength{\tabcolsep}{8pt}
\begin{tabular}{| r|rrr |} 
\hline\hline 
$k$ &   $r_k^\text{veto}$ &   $r_k^\text{can}$ &  $r_k^\text{thr}$\\ [0.5ex] 
\hline 
1 & 0.3 & 0 & 0.3 \\ 
2 & 0.25 & 0.07 & 0.25\\
3 &  0.87 & 0.33 & 0.47\\
4 & 2.78 & 1.04 & 2.48\\
5 & 6.30 & 3.25 & 6.30\\
6 &  & 7.94 & \\
7 &  & 17.40 & \\
8 &  & 36.63 & \\
9 &  & 75.28 & \\
10 &  & 152.52 & \\
11 &  & 306.61 & \\
14 &  & 2464.29 & \\%[1ex]
\hline
\end{tabular}

\end{table}

In order to evaluate applicability of the veto function in natural systems, we
analyze the functions in a Boolean network based on a real living system. The
data set is a collection of biological input-response of 152 nodes of
intracellular signal transduction network in the form of Boolean truth tables
each of which corresponds to the update function of a node
\cite{Helikar:2008,Rue:2010}.
We investigate the over-representation of veto functions, threshold functions
and canalizing functions in this system.

For each property $\pi$, we count the $k$-ary functions in the data set as:
\begin{equation}
a_k^{(\pi)} = |\{ i \in \{1,2,\dots,152\} : f_i \text{ is }k\text{-ary with property }\pi \}|~.
\end{equation}
Table~\ref{tab:raw} provides an overview of these counts.

Then $a_k^{(\pi)} / a_k$ is the fraction of these functions, $a_k$ being the total count of $k$-ary functions in the data set.
In order to quantify the preference of property $\pi$, we compare this fraction to a null model of uniformly drawing $k$-ary functions
without spurious inputs. Under this null model, the expected fraction of functions with property $\pi$ is
$|\B_k^\ast,(\pi)| / |\B_k^\ast|$. We call the {\em over-representation} of property $\pi$ at arity $k$, the logarithm of the
ratio between the observed fraction and that expected under the null model, so
\begin{equation}
r_k^{(\pi)}  = \log \left(\frac{a_k^{(\pi)}}{a_k}\right) - \log \left(\frac{|\B_k^\ast,(\pi)|}{|\B_k^\ast|}\right)
\end{equation}
Table~\ref{tab:over} lists these values. The over-representation of veto functions is at least 
as large as that of threshold functions and strictly larger than that of canalyzing functions,
considering the value of $r_k^{(\pi)}$ for arities $k \le 5$ where all these types of functions are present.

%%%%%%%%%%%%%%%%%%%%%%%%%%%%%%%%%%%%%%%%%%%%%%%%%%%%%%%%%%%%%%%%%%%%%
\section{Cell cycle networks} \label{sec:cellcyc}
%%%%%%%%%%%%%%%%%%%%%%%%%%%%%%%%%%%%%%%%%%%%%%%%%%%%%%%%%%%%%%%%%%%%%

\begin{figure}
\centering
\includegraphics[width=0.46\textwidth]{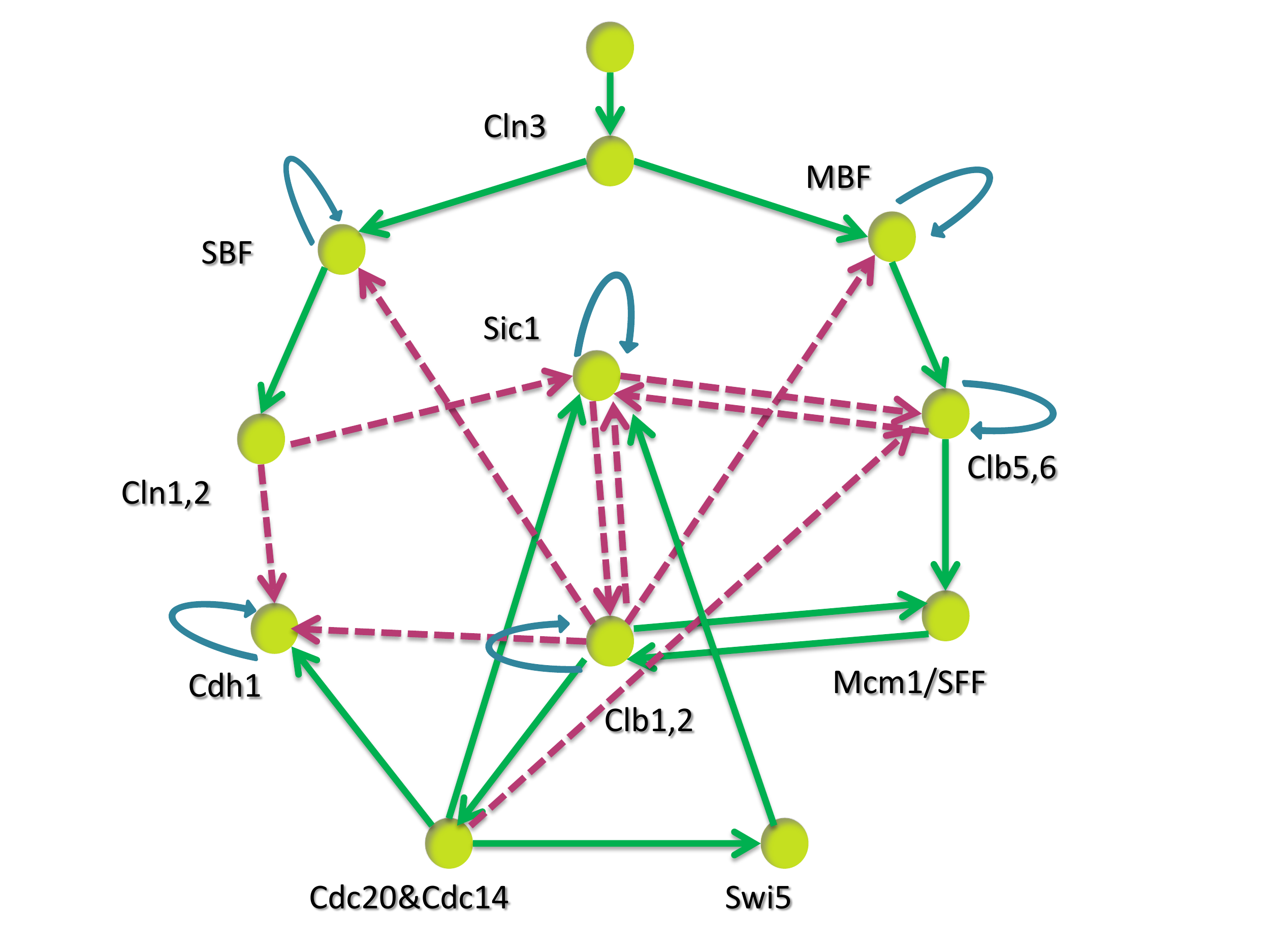}
\caption{\label{fig:bud}
(Color online) Functional network using veto functions for the budding yeast cell cycle. Solid arrows represent activators,
dashed arrows inhibtors. Departing from the wild type network based on threshold functions \cite{Rybarsch:2012}, the depicted
network is obtained by deleting six interactions:
the two activations from
Mcm1/SFF to Swi5 and to Cdc20$\&$Cdc14; and the four inhibitions from Clb5,6 to Cdh1, from Cdh1 to Clb1,2, from Cdc20$\&$Cdc14 to
Clb1,2, and from Clb1,2 to Swi5.
}
\end{figure}

\begin{figure}
\centering
\includegraphics[width=0.46\textwidth]{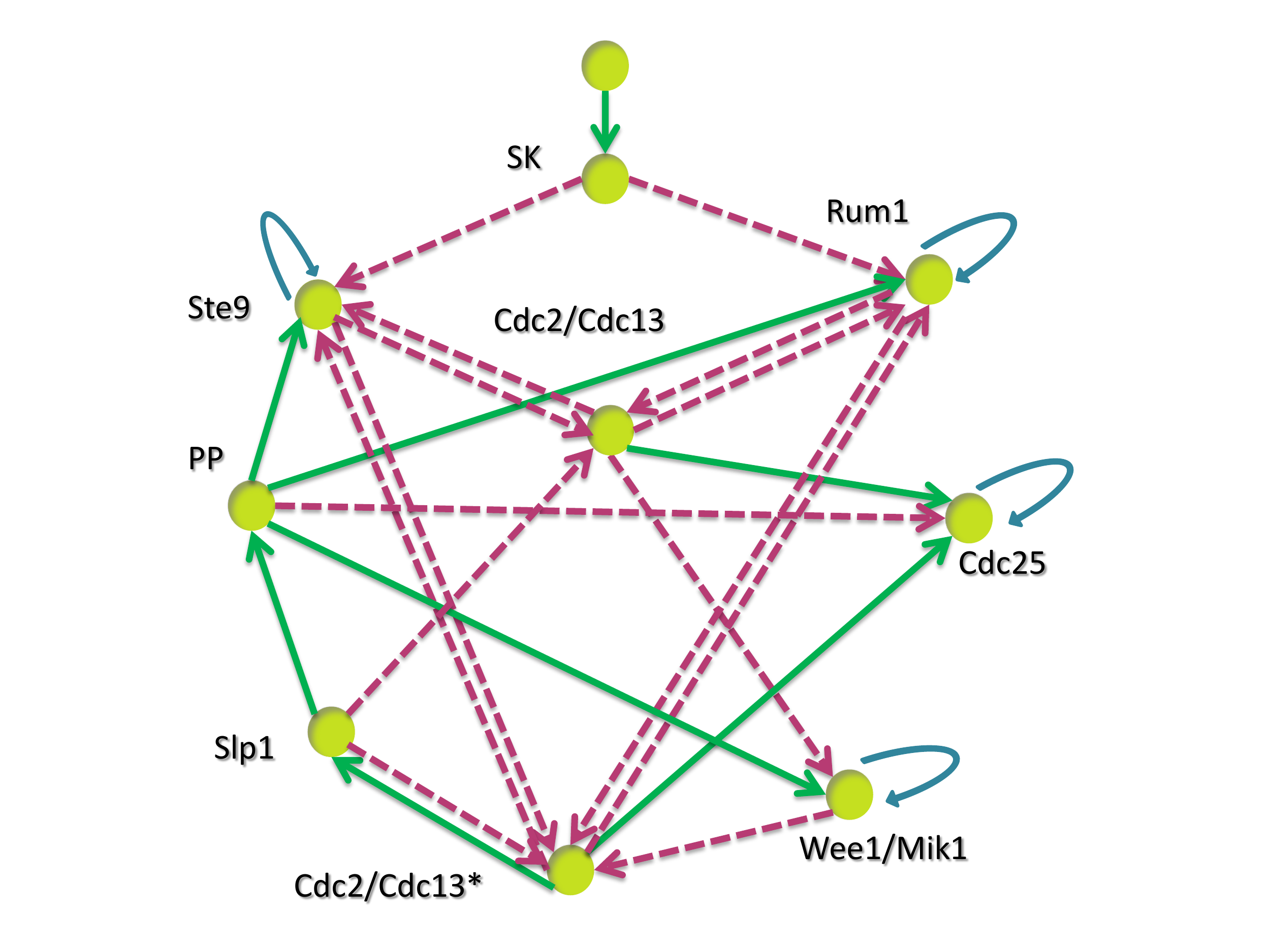}
\caption{\label{fig:fiss}
(Color online) Functional network using veto functions for the fission yeast cell cycle. Solid arrows represent activators,
dashed arrows inhibtors. The wiring is identical to the one given by Rybarsch and Bornholdt \cite{Rybarsch:2012}, bottom of Figure 5 there.
}
\end{figure}

Cell division has been one of the first biological processes to be described in terms of Boolean networks, using with few (around 10) nodes
\cite{Li:2004,Lau:2007,Davidich:2008,Rybarsch:2012,Boldhaus:2010}. In the Boolean discretization, the cell cycle is a sequence of state vectors $\sigma(0),\sigma(1),\dots,\sigma(T)\in\{0,1\}^N$ where $\sigma_j(t)$ indicates the presence or absence
of molecular species $j$ at time step $t$. A Boolean network on $N$ nodes is called {\em functional} if it generates the cell cycle sequence given $\sigma(0)$ as an initial condition.
Most earlier approaches describe functional networks using threshold functions. Here we investigate functional networks using veto functions.

In a functional network, each node $i$ independently fulfills the input-output mapping given by the sequence. Thus the problem of finding {\em all} functional networks is fully solved by independently
finding the set $S_i^\text{veto}$ of functions generating this mapping \cite{Lau:2007}.
\begin{equation}\label{eq:ve}
S_i^\text{veto}= \{f \in \B^\text{veto}_N | \forall t\in \{1,\dots,T\}:f(\sigma(t-1))=\sigma_i(t)\}~.
\end{equation}
with $\B^\text{veto}_N$  denoting the set of all $N$-ary veto functions (cf.\ section \ref{sec:counting}). Since the solutions at each node $i$ combine independently, the number of functional networks based on veto functions
is
\begin{equation}
H^\text{veto} = \prod_{i=1}^N |S_i^\text{veto}|.
\end{equation}

For the cell cycle of the species {\em S.\ cerevisiae} (budding yeast, $N=11$) \cite{Li:2004}, we compute $H^\text{veto} = 1.15 \times 10^{28}$, to be compared to $H^\text{thr}=1.6 \times 10^{33}$ functional networks using threshold functions.
Fig.~\ref{fig:bud} shows one of the functional  networks with veto functions. It has been selected such that the wiring is closest to the so-called wild type \cite{Li:2004,Rybarsch:2012} based on interactions with evidence in the literature. Departing
from the wiring of the wild-type, the network in Fig.~\ref{fig:bud} is obtained by deleting six interactions, see the caption for details.

For the cell cycle of the species {\em S.\ pombe} (fission yeast, $N=10$) \cite{Davidich:2008}, we compute $H^\text{veto} = 2.97 \times 10^{27}$, to be compared to $H^\text{thr}=2.4 \times 10^{27}$ functional networks using threshold functions.
Fig.~\ref{fig:fiss} shows one of the functional  networks with veto functions. The wiring is identical to the wild type network using threshold functions \cite{Rybarsch:2012}. We remark that node Cdc2/Cdc13 is treated different from the other nodes.
This node does not have an activating connection in the wild type wiring. Here we use a varied type of veto function where a node is active in the absence of inhibiting inputs (even though there is no activating input). This is analogous to the treatment
with threshold functions where a negative threshold is assigned to Cdc2/Cdc13.

%%%%%%%%%%%%%%%%%%%%%%%%%%%%%%%%%%%%%%%%%%%%%%%%%%%%%%%%%%%%%%%%%%%%%
\section{Sensitivity} \label{sec:sensitivity}
%%%%%%%%%%%%%%%%%%%%%%%%%%%%%%%%%%%%%%%%%%%%%%%%%%%%%%%%%%%%%%%%%%%%%

The tendency of a Boolean function  to change output value in response to a
changing input is quantified by the activity \cite{Shmulevich:2004},
defined as
\begin{equation}\label{eq:activity}
\alpha_{l}(f) = 2^{-k} \sum_{\sigma \in \{0,1\}^k}  \partial^{(l)} f(\sigma),
\end{equation}
with $\partial$ as defined in Eq.~(\ref{eq:derivative}).
Thus the activity is the probability that a
perturbation (negation of state) at input $l$ propagates to the output of the
function when all other inputs are kept fixed.

In analogy to the usual partial derivative, $\partial^{(i)} f$ indicates whether
a change in the $i$-th input variable causes a change of output. For an input
vector $\sigma \in \{0,1\}^k$,
\begin{equation}\label{eq:derivative}
\partial^{(i)} f (\sigma) = \left\{\begin{array}{cl}
1 & \textrm{if } f (\sigma) \neq f (\sigma^{\updownarrow i})\\
0 & \textrm{otherwise}
\end{array}\right.
\end{equation}
with $i \in \{1,\dots,k\}$ and $^{\updownarrow i}$ indicating negation at the $i$-th component (Eq.~\eqref{eq:negation}).

Let us consider a $k$-ary veto function $f$.
As before, we denote the activating inputs by $A$, inhibiting inputs by $I$. Let  $m:=|I|=m$
and assume absence of spurious inputs, so $|A|=k-m$. Let us consider the set $X_l$ of state vectors
where flipping the state of the $l$-th component causes $f$ to change output,
\begin{equation}
X_l = \{ \sigma \in \{0,1\}^k : f(\sigma^{\updownarrow l}) \neq f(\sigma) \}
\end{equation}
so $\alpha_l(f) = 2^{-k} |X_l|$.

An inhibiting input $l$ switches the output if and only if there is at least one activation and all other inhibitors are off,
\begin{equation}
X_l = \{ \sigma \in \{0,1\}^k : (\forall i_I\setminus\{l\} : \sigma_i=0) \wedge \exists j \in A: \sigma_j =1 \}
\end{equation}
This comprises $|X_l| = 2 (2^{k-m} -1)$ state vectors, so the activity of an inhibitor is
\begin{equation}
 \alpha_{l}(f) = \frac{2 (2^{k-m}-1)}{2^{k}}~.
\end{equation}

When switching the state at an activating input $l \in A$, the output of $f$ changes if and only if all other inputs
are off,
\begin{equation}
% \partial^{(l)}f\neq 0  \Leftrightarrow \forall i_{\neq l} \in I\cup A: \sigma_{i}=0~.
X_l = \{ \sigma \in \{0,1\}^k : \forall i \in \{1,\dots,k\}\setminus\{l\} : \sigma_i=0 \}~.
\end{equation}  
Here we have $|X_l| =2$ state vectors only. The activity of an activator is
\begin{equation} 
 \alpha_{l}(f) = \frac{2}{2^{k}}
\end{equation}

The sensitivity is the sum of activities of all inputs 
\begin{equation} \label{eq:sensitivity}
s(f) = \sum_{i=1}^k \alpha_{i}~.
\end{equation}
For the veto functions with $m$ inhibitors and $k-m$ activators, we obtain
\begin{equation}\label{eq:veto_sens_km}
s(f) = \frac{m(2^{k-m}-2)+k}{2^{k-1}}~.
\end{equation}

\begin{table}
\caption{Characterization of veto functions by sensitivity $s$. Depending on the total number of inputs $k$ and the
number $m$ of inhibitors among these, functions would lead to frozen ($s<1$), critical ($s=1$), chaotic ($s>1$)
dynamics in the annealed approximation. Only functions without irrelevant (spurious) inputs are considered, so the
number of activators is $k-m$.} 
\begin{tabular}{ | c || c |c | c| c | c | c |}
\hline\hline 
$k$ &   $m=0$ &   $m=1$ &     $m=2$  &  $m=3$ & $m=4$ & $m>4$\\ [0.5ex] 
\hline 
  1 & critical  & frozen  & --- & ---& ---&---  \\ \hline
  2 & critical  & critical  & frozen  &  ---& ---& --- \\ \hline
  3 & frozen  & chaotic  & chaotic & frozen & --- & --- \\ \hline
  4 & frozen  & chaotic  & critical  & frozen & frozen & --- \\ \hline
   $ >4$ & frozen  & chaotic  & chaotic  & frozen & frozen & frozen \\
  \hline  
\end{tabular}
\label{table:phase-veto} 
\end{table}

The sensitivity is the crucial parameter in the annealed approximation \cite{Derrida:1986,Shmulevich:2004}.
It predicts a transition from ordered (convergent) to chaotic (divergent) dynamics at a sensitivity value 1 in large networks.
For networks homogeneously built with veto functions of $m$ inhibitors and $k-m$ activators, the expected
dynamic phase is obtained by evaluating Eq.~(\ref{eq:veto_sens_km}) and listed in Table~\ref{table:phase-veto}.
Non-frozen dynamics is rarely obtained. For $k\ge 4$, only $m=1$ or $m=2$ lead to $s>1$, otherwise $s<1$.

\begin{figure}
\centering
\includegraphics[width=0.46\textwidth]{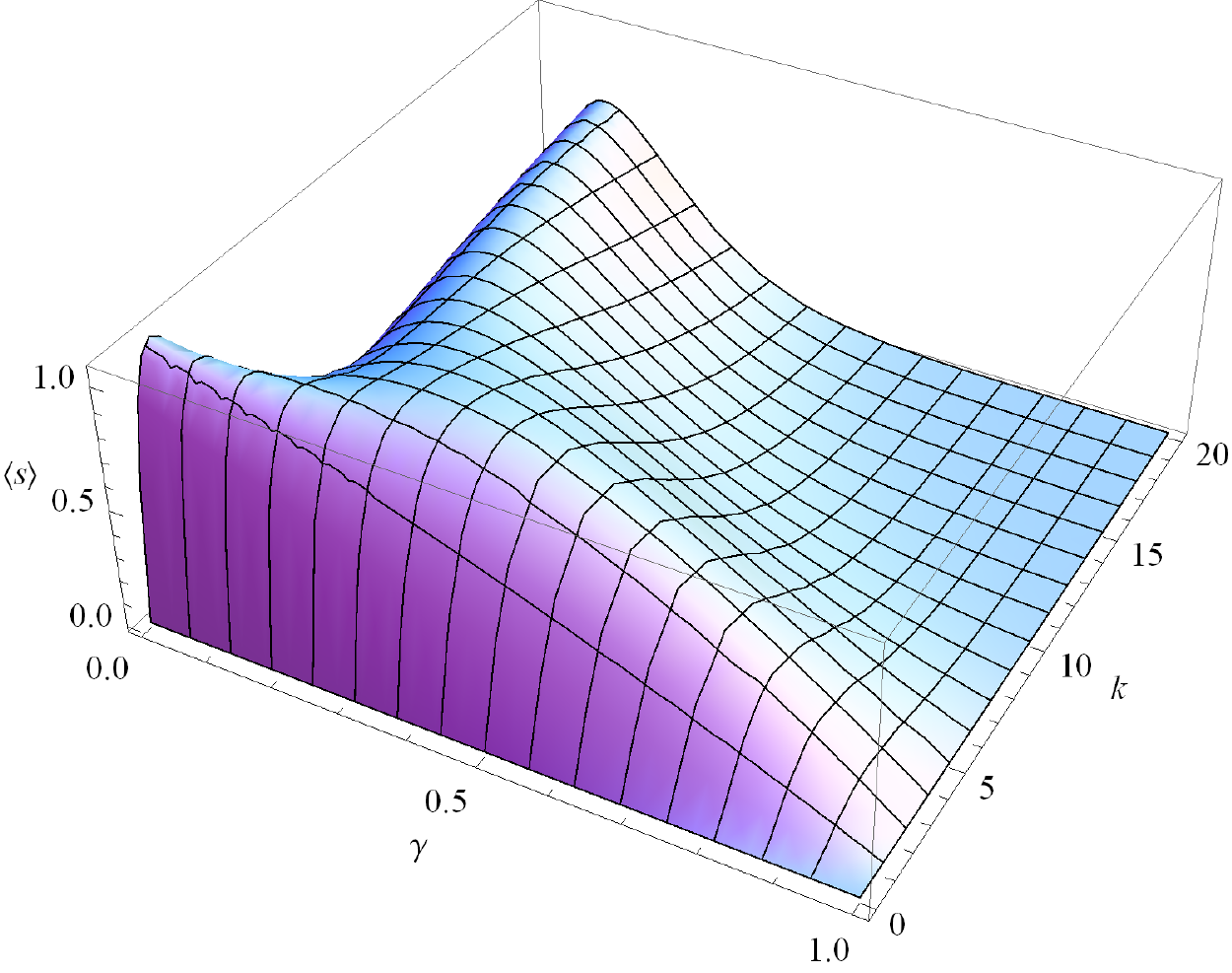}
\caption{\label{fig:av-sen}
(Color online) Average sensitivity according to Equation~(\ref{eq:av-sen}) for an ensemble of veto functions with inhibitor probability $\gamma$.
The number of inputs $k$ is a fixed integer parameter of the ensemble (no averaging over $k$).}
\end{figure}

Now let us consider a statistical ensemble of veto functions with $k$ inputs where the number of inhibitors $m$ is distributed
binomially with parameter $\gamma$. Thus in generating a function, we decide for each of the $k$ inputs independently, if it
is taken as an inhibitor (with probability $\gamma$) or an activator (with probability $1-\gamma$). Then the ensemble averaged
sensitivity is
\begin{equation} \label{eq:av-sen}
\langle s \rangle = \sum_{m=0}^k {k \choose m} \gamma^m (1-\gamma)^{k-m} \left[\frac{m(2^{k-m}-2)+k}{2^{k-1}}\right]~.
\end{equation}
Using
\begin{equation}
\gamma^m (1-\gamma)^{k-m} 2^{k-m} = (2-\gamma)^k \left(\frac{\gamma}{2-\gamma}\right)^m \left(1-\frac{\gamma}{2-\gamma}\right)^{k-m}
\end{equation}
the $m2^{k-m}$ effectively sums as a binomimal distribution with parameter $\gamma/(2-\gamma)$. We arrive at:
\begin{align}
\langle s \rangle = \frac{1}{2^{k-1}}[(2-\gamma)^{k-1} k \gamma - 2k \gamma + k]\\
= \frac{k}{2^{k-1}}[(2-\gamma)^{k-1} \gamma - 2 \gamma + 1]
\end{align}
This ensemble averaged sensitivity is plotted in
Fig.~\ref{fig:av-sen}. These values $ {\langle}s{\rangle}$ never exceed $1$.
In contrast to concrete choices $(k,m)$, cf.\ Table~\ref{table:phase-veto},
the ensemble of independent stochastic assignment of
inhibitors and activators to veto functions always gives ordered dynamics.
Statistical ensembles sufficiently concentrated at functions with
$m=1$ inhibitors yield an average sensitivity above $1$. The binomial
distribution of the number of inhibitors, however, is sufficiently broad to
ensure that contributions from functions with low sensitivity dominate.

%%%%%%%%%%%%%%%%%%%%%%%%%%%%%%%%%%%%%%%%%%%%%%%%%%%%%%%%%%%%%%%%%%%%%
\section{Closing remarks and outlook}
%%%%%%%%%%%%%%%%%%%%%%%%%%%%%%%%%%%%%%%%%%%%%%%%%%%%%%%%%%%%%%%%%%%%%

The idea of strong inhibitory inputs of veto type has been used in models of
neurons before. It dates back at least to the work by McCulloch and Pitts
\cite{McCulloch:1943} where the veto of an inhibitor is made explicit as a rule
of the model:
``The activity of any inhibitory synapse absolutely  prevents excitation of the
neuron at that time.''
The idea of a veto is taken further by electrochemically detailed models of
neural dynamics; there it combines with a spatial aspect where an inhibitor
exerts a local veto, suppressing the effect only of activating signals
\cite{Koch:1983}. The spatial aspect owes to the fact that neural networks are
embedded in real physical space. Nevertheless also in the context of
regulatory networks, we may generalize veto functions such that each inhibitor
vetos only against a subset of all activators.

Strong inhibition is also used in studies of cell cycle
networks in a model with graded (non-Boolean) response by Burda, Zagorski and
co-authors \cite{Burda:2011,Zagorski:2013}. Networks with veto functions are
obtained when discretizing the response functions used in their model. Veto
rules appear in further, also non-biological contexts. In networks of detectors
for gravitational waves, some of the devices may be given a veto function in
order to suppress false positive signals \cite{Wen:2005}.

Classes of functions and their parametrizations are to be explored
further, under comparison with empirical data sets. This will lead to more and
refined null models being able to separate global effects from network architecture
from local ones given by the use of logical functions with particular
properties. 

%%%%%%%%%%%%%%%%%%%%%%%%%%%%%%%%%%%%%%%%%%%%%%%%%%%%%%%%%%%%%%%%%%%%%
\section{Acknowledgments}
%%%%%%%%%%%%%%%%%%%%%%%%%%%%%%%%%%%%%%%%%%%%%%%%%%%%%%%%%%%%%%%%%%%%%

The authors acknowledge helpful comments from Johannes Rauh (MPI-MiS Leipzig),
This work has been funded by VolkswagenStiftung under the call
Complex Networks as Phenomena across Disciplines. K.~K.\ acknowledges
partial funding by FWF (Austria) through
project SFB F43, RNA regulation of the transcriptome.

\bibliography{vetofunc}

\end{document}